\documentclass[aps]{revtex4}

\begin{document}

\title{Large $N_c$ QCD and Models of Exotic Baryons}

\author{Thomas D.~Cohen\,\footnote{\,\uppercase{W}ork supported in part by
the \uppercase{U}.\uppercase{S}. \uppercase{D}epartment of
\uppercase{E}nergy under grant  \uppercase{DE}-
\uppercase{FG}02-93 \uppercase{ER}-40762.}}

\affiliation{Department of Physics, University of Maryland \\
College Park, MD 20742 USA\\
E-mail: cohen@physics.umd.edu}

\begin {abstract} Exotic baryons have been predicted in the context of
the chiral soliton model. These states have been identified with
strangeness +1 resonances reported in a number of experiments. In
this talk it is pointed out that the technique used to quantize
these solitons in most conventional treatments depends on
dynamical assumptions beyond those in standard large $N_c$
physics.  These additional assumptions have never been
justified.\end{abstract}

\maketitle

2003 saw the first report of a narrow resonant baryon state with
strangeness +1 and a mass of approximately 1.54 GeV\cite{exp};
this state has been denoted the $\theta^+$.  This claim
stimulated an extraordinarily high level of both experimental and
theoretical research.  The reason for this intense interest stems
largely from the fact that such a state is manifestly exotic: it
cannot be made as a three quark state.  The minimum number of
quarks to construct a state with these quantum numbers is 5 (four
quarks and one anti-quark) and accordingly such states are often
referred to as pentaquarks.  I will not review the experimental
situation here except to note that it is quite confused: numerous
experiments appear to confirm the existence of a narrow exotic
state at a similar mass while many others see no indication of
it.  While it is extremely important that this experimental
situation be sorted out---as will presumably happen in the next
couple of years---the issues of relevance here do not strongly
depend on how things are ultimately resolved experimentally.
Instead, this talk focuses on the theory side and, in particular,
on how large $N_c$ QCD can help understand the situation.  The
work discussed here summarizes work in refs. \cite{Coh1,Coh2} and
for a more complete discussion the reader is referred to the
original work.

Most theories concerning pentaquarks is based on modeling QCD
rather than QCD itself. Of course, the number of models is quite
large and I will not attempt to review the full spectrum of
models. Instead, I will focus on one class of models---the chiral
soliton models---since they have a rather special status. In the
first place, a chiral soliton model\cite{DPP} was used to predict
a narrow pentaquark state with virtually the same mass as was
ultimately reported experimentally.  In contrast, most other
models were postdictions with parameters fit with a knowledge of
the mass.  Clearly, true predictions carry  much greater weight:
as has been said,``Predictions are difficult---especially about
the future''.  (It is interesting to note that an internet search
to find the origin of this quote led to a site attributing it to
Yogi Berra---the former New York Yankee catcher---and another
site attributing it to Neils Bohr---the Danish physicist.) Perhaps
more importantly, the prediction was quite insensitive to the
details of the model; the functional form of the soliton profile
function played no role. All that mattered dynamically was the
structure of the model, the values of SU(3) breaking parameters
(fixed in the non-exotic sector) and the identification of the
nucleon state of the multiplet.

From a theory perspective this model insensitivity is crucial. It
has long been known that there are predictions of soliton models
for which the details of the soliton model are totally irrelevant
(such as the ratio  $g_{\pi NN}/g_{\pi N \Delta}$ \cite{AN}). {\it
All} known cases where predictions of soliton models do not
depend on the dynamical details are also known to be true in
large $N_c$ QCD. This can be demonstrated by the use of large
$N_c$ consistency rules\cite{cons}.  This suggests that the
predicted $\theta^+$ properties might also be model-independent
predictions of large $N_c$ QCD.  If true, this would be really
important: we would then essentially understand the structure of
the $\theta^+$ modulo SU(3) symmetry breaking effects and
higher-order 1/$N_c$ corrections. Unfortunately, however, this
simply is not the case.  The apparent insensitivity to the model
details arises from an inconsistent treatment of the quantization
of the soliton models in the sense of large $N_c$ counting.

To understand why this is so, we need to review the standard
methods for quantizing solitons. To begin, consider a topological
soliton model which has SU(2) flavor symmetry and is based on a
nonlinear sigma model. The chiral fields are given in terms of a
matrix $U=\exp \left ( i \frac{\vec{\tau} \cdot \vec{\pi} }{f_\pi}
\right )$ where $\vec{\pi}$ are the pions.  Large $N_c$ QCD
justifies solving the theory classically and the lowest energy
solution having winding number unity (corresponding to baryon
number unity) is a ``hedgehog'' of the form $U_0=\exp \left ( i
{\vec{\tau} \cdot \hat{r} f(r) } \right )$ with
$f(0)-f(\infty)=\pi$.  Note that such a configuration correlates
isospace and ordinary space and thereby breaks both rotational
and isospin symmetries. Accordingly such a classical
configuration cannot correspond directly to a single physical
state.  Rather, the hedgehog corresponds to a band on nearly
degenerate states---in much the same way as a deformed
Hartree-Fock solution in nuclear physics corresponds to a band of
low-lying states.

To obtain physical states one needs a method to project from the
hedgehog onto states with physical quantum numbers.  Adkins,
Nappi and Witten (ANW) introduced a semiclassical  method
(justified at large $N_c$) for doing this \cite{ANW}.  One
introduces an ansatz for a  time-dependent $U$ given by
$U(\vec{r},t) = A^\dagger(t) U_0(\vec{r}) A(t)$, where $A(t)$ is a
time-dependent, space-independent, SU(2) matrix. Inserting this
into the lagrangian density of the model and integrating over
space gives a collective lagrangian which only depends on $A$ and
$\partial_t A$.  This can be Legendre transformed into a
collective hamiltonian:
\begin{equation}
H = M_0 + \frac{1}{2{ I}} \left (\frac{\partial^2}{\partial
a_0^2} +\frac{\partial^2}{\partial a_1^2} +
\frac{\partial^2}{\partial a_2^2} + \frac{\partial^2}{\partial
a_2^2} \right )
\end{equation}
with the constraint $|a_0|^2+|a_1|^2+|a_2|^2+|a_2|^2 =1$.  The key
to what follows is the $N_c$ scaling: $M_0\sim N_c$ and ${ I}\sim
N_c$. This corresponds to a slowly rotating hedgehog at large
$N_c$ and its motion can be separately quantized:
\begin{equation}
M_{I=J} = M_0 + \frac{J(J+1)}{2 { I}}
\end{equation}
Note all states have $I=J$ and one can consistently quantize the
system to have I=J being half integers. The lowest two states have
I=J=1/2 (nucleon) and $I=J=3/2$ ($\Delta$) and thus $M_\Delta -
M_N= 3/I \sim 1/N_c$.  The collective wave functions  are just the
Wigner D matrices (appropriately normalized).

It should be stressed that this approach is based on an ansatz
and as such one must check the self-consistency of the approach.
Ultimately it is justified at large $N_c$ since the collective
motion is adiabatic: angular velocities go like $J/{ I} \sim
N_c^{-1}$ and are truly collective, covering the full angular
space (of order $N_c^0$) yielding frequencies (and hence
excitation energies of order $N_c^{-1}$.  This, in turn, implies a
Born-Oppenheimer separation of the slow collective motion from the
faster modes associated with vibrations of the meson fields (with
time scales of order $N_c^0$ and hence energies of order
$N_c^0$).  Because of this scale separation the collective modes
can be quantized separately from the intrinsic vibrations.

Now let us turn to the problem of relevance to $\theta^+$
physics: SU(3) solitons.  Shortly after the ANW quantization was
introduced it was extended by a number of workers to solitons in
models with SU(3) flavor\cite{SU3Quant}.  Extending the models
from SU(2) to SU(3) flavor is intellectually straightforward. The
only essential new feature is the inclusion of a topological
Wess-Zumino-Witten term which builds in the anomalies\cite{Wit1}.
For simplicity, first consider the case of exact SU(3) symmetry.
The standard semiclassical  approach is then to first solve the
problem classically which yields a hedgehog configuration in a
two-flavor subspace (which we will take to an intrinsic u-d
subspace by convention).  Following the ANW ansatz one introduces
a time-dependent, space-independent, global SU(3) rotation $A(t)$.
At this stage the parallel of the two flavor case is virtually
exact.  However, the Wess-Zumino-Witten term introduces a
constraint: In the body-fixed (co-rotating) frame the hypercharge
must be $N_c/3$.  This constraint plays a critical role in what
follows.

Going through this procedure yields a collective rotation
hamiltonian of the form:
\begin{equation}H_{\rm rot} = M_0+
\frac{1}{2 I_1} \sum_{A=1}^3 {\hat{J}_A'}{}^2 \, + \, \frac{1}{2
I_2} \sum_{A=4}^7 {\hat{J}_A'}{}^2   \; , \label{collective}
\end{equation}
where the prime indicates the generator as measured in a
co-rotating frame, and $I_1$ ($I_2$) is the moment of inertia for
motion within (out of) the original SU(2) subspace.  Note that
there is kinetic energy in the intrinsic 8 direction as it leaves
the original hedgehog unchanged.  The physics associated with
this direction is encoded in the constraint.  As noted by
Witten\cite{Wit2} this is quite analogous to the problem of a
charge particle moving in the field of a magnetic monopole.

This procedure yields masses given by
\begin{eqnarray}
M & = &  M_0 + \frac{C_2}{2 I_2} + \frac{(I_2 - I_1) J (J+1) }{2
I_1 I_2} - \frac{N_c^2}{24 I_2} \; , \nonumber
\label{mass} \\
{\rm with} \; \; &C_2 & =  \left( p^2 + q^2 + p q + 3(p +q)\right
)/3  \; .
 \end{eqnarray}
$C_2$ is  the quadratic Casimir, and is labeled  by $p,q$ which
specify the SU(3) representation.  The constraint due to the
Wess-Zumino-Witten term imposes the restriction that the
representation must have a state with $Y=N_c/3$ and implies that
angular momentum is determined by the condition that $2J+1$ equals
the number of states with S=0.  Plugging in $N_c=3$, one sees the
lowest representations are (p,q)=(1,1) (spin 1/2 octet),
(p,q)=(3,0) (spin 3/2 decuplet) and (p,q)=(0,3) (spin 1/2
anti-decuplet).  This last representation is clearly exotic.

Diakonov, Petrov and Polyakov proposed to take these
anti-decuplet states seriously\cite{DPP}.  The predicted value
depends  on $I_2$ which was fit by identifying the N(1710) as a
member of the multiplet and by using SU(3) breaking effects
included perturbatively with all parameters determined in the
non-exotic sector. The fact the width of the state came out as
relatively small was taken as a self-consistent justification for
not treating the $\theta^+$ as a large $N_c$ artifact.

The question which needs to be addressed is whether the
rigid-rotor type semiclassical  projection used here is kosher for
the exotic states.  Since the question ultimately comes down to
whether the Born-Oppenheimer separation is justified at large
$N_c$ for these states, care should be taken to keep $N_c$
arbitrary and large throughout the analysis.  In particular one
ought not set $N_c=3$ when imposing the constraint of the
Witten-Wess-Zumino term, at least when doing formal studies of the
$N_c$ dependence.  In doing this one sees that the flavor SU(3)
representations at large $N_c$ differ from their $N_c=3$
counterparts.  The lowest-lying representation consistent with the
Witten-Wess-Zumino term constraint has $ \left (p,q \right ) =
\left( 1, \frac{N_c-1}{2} \right) $ and has $J=1/2$.  Thus, it is
a clear analog of the octet representation and will be denoted as
the ``8'' representation. (The quotes are to remind us that it is
{\it not} in fact an octet representation.)  Similarly the next
lowest representation,  has $ \left ( p,q \right ) = \left(
3,\frac{N_c-3}{2} \right ) $ and has $J=3/2$.  It is the large
$N_c$ analog of the decuplet and will be denoted ``10''. The
salient feature of the anti-decuplet is that its lowest
representation contains an exotic S=+1 state.  Thus its large
$N_c$ analog is the lowest representation containing manifestly
exotic states.  This representation is $ \left ( p,q \right ) =
\left( 0, \frac{N_c+3}{2} \right ) $ and has $J=1/2$; it will be
denoted ``$\overline{10}$''.

Now let us look at the excitation energy of the exotic states
which can be computed from eq.~(\ref{mass}):
\begin{equation}
 M_{``\overline{10}"} - M_{``8"}  = \frac{3 + N_c}{4 I_2} \; .
 \label{10bar8quote}
\end{equation}
Noting that $I_2 \sim N_c$, one sees that this implies that the
excitation energy of the exotic state is of order $N_c^0$.  This
may be contrasted to the excitation of the non-exotic ``10''
representation which is of order $1/N_c$.

The fact the standard semiclassical quantization gave excitation
energies of order $N_c^0$ for exotic states means that the
approach is not justified for such states.  Recall that the
analysis was justified self-consistently via a Born-Oppenheimer
scale separation.  For the non-exotic states this is justified.
However, for the exotic states the characteristic time associated
with the excitations (one over the energy difference) is order
$N_c^0$. This is the same characteristic time as the ``fast''
vibrational excitation of the meson fields.  One cannot therefore
justify treating the ``collective'' degrees of freedom
separately.  Thus the prediction of $\theta^+$ properties via
this collective quantization procedure cannot be justified from
large $N_c$.

In fact, there are many other ways to see that approach is not
justified by large $N_c$.  There is an extensive discussion of
these in ref. \cite{Coh2}.  Here we will briefly mention a couple
of these.   As discussed above, ref.~\cite{DPP} stressed the small
numerical value of the width to justify the treatment
self-consistently. However, from the perspective of formal large
$N_c$ consistency, this numerical value is essentially irrelevant.
The key question is how does the width depend parametrically  on
$N_c$?  If the standard collective quantization procedure outlined
above had been justified by large $N_c$ physics, then it should
become exact in the large $N_c$ limit in the sense of giving an
exact value for the mass. If this is not the case then some kind
of {\it ad hoc} correction must be added on and there is no {\it
a priori} reason for it to be small unless it vanishes at large
$N_c$. This in turn means the width must go to zero at large
$N_c$. One way to see this is simply that if the width is
non-zero, then an asymptotic state doesn't exist and the concept
of an exact mass becomes ill defined. Alternatively one can view
the width as originating from an imaginary contribution to the
mass. However, the standard collective quantization gives a real
value, and had it become exact at large $N_c$ one would have zero
widths in this limit.

The upshot of all this is that if the approach is justified, then
at a formal level the width must approach zero at large $N_c$.  Of
course, for non-exotic states such as the decuplet, this is true.
The reason is simply phase space.  As $N_c$ grows, the excitation
energy for these states drops thereby killing the phase space for
decay.  In contrast, as shown recently by  Praszalowicz\cite{Pra}
the width of the $\theta^+$ as calculated via the standard
collective approach is order $N_c^0$.  This demonstrates that the
procedure is not self consistent.

Another perspective is given by the spin-flavor contracted SU(6)
symmetry which can be derived on general grounds from QCD using
large $N_c$ consistency rules \cite{cons}.  It is precisely the
existence of such a symmetry that requires predictions independent
of details in soliton models to arise as true large $N_c$
model-independent results of QCD.  However, for three flavors the
results of such an analysis are quite well known. One predicts
exactly the states in a large $N_c$ naive quark model---exotic
collective states are not obtained in this model-independent
approach.  Thus, {\it a priori} there is no fundamental reason to
believe that the quantization procedure discussed above which
gives rise to collective exotic states is consistent with QCD. In
contrast, the nonexotic states clearly are.

For another perspective on problems with the standard method used
to quantize these solitons see Igor Klebanov's talk in this
volume and his work with collaborators at Princeton in ref.
\cite{Kleb}.

While there appears to be compelling evidence that the approach
fails there is an obvious question as to why.  Of course, at
certain level there is no mystery.  The approach is based on an
ansatz and the ansatz needs to be shown to be self consistent.
While the scales are such that the properties of non-exotic states
can be self-consistently described, they are also such that the
self-consistency fails for the exotic states.  At a deeper level
the failure is due to the mixing of intrinsic vibrational  modes
with the collective rotational modes.  For systems without
velocity dependent forces such modes are orthogonal and hence
cannot mix at leading order.  However, as discussed in
ref.~\cite{Coh2}, velocity-dependent forces spoil this
orthogonality and allow mixing.  The Witten-Wess-Zumino term
gives rise to precisely such velocity-dependent interactions and
spoils the orthogonality.

In conclusion, the standard collective quantization method as
applied to exotic states such as the $\theta^+$ cannot be
justified from large $N_c$ QCD.  It is logically possible, of
course, that such a procedure can be justified for some other
reason, but at present no such justification is known.  Thus one
should view any predictions based on this procedure with real
caution.

\end{document}